# The Extreme Space Weather Event in February/March 1941

Hisashi Hayakawa* (1 – 4), Sean P. Blake (5 – 6), Ankush Bhaskar (5 – 6), Kentaro Hattori (7), Denny M. Oliveira (5, 8), Yusuke Ebihara (9 – 10)

(1) Institute for Space-Earth Environmental Research, Nagoya University, Nagoya, 4648601, Japan

(2) Institute for Advanced Researches, Nagoya University, Nagoya, 4648601, Japan

(3) UK Solar System Data Centre, Space Physics and Operations Division, RAL Space, Science and Technology Facilities Council, Rutherford Appleton Laboratory, Harwell Oxford, Didcot, Oxfordshire, OX11 0QX, UK

(4) Nishina Centre, Riken, Wako, 3510198, Japan.

(5) Heliophysics Science Division, NASA Goddard Space Flight Center, Greenbelt, MD, USA

(6) Catholic University of America, Washington DC, United States

(7) Graduate School of Science, Kyoto University, Kyoto, 6068501, Japan.

(8) Goddard Planetary Heliophysics Institute, University of Maryland, Baltimore County, Baltimore, MD, United States

(9) Research Institute for Sustainable Humanosphere, Kyoto University, Uji, 6110011, Japan

(10) Unit of Synergetic Studies for Space, Kyoto University, Kyoto, 6068306, Japan

* hisashi@nagoya-u.jp

**Abstract**

Given the infrequency of extreme geomagnetic storms, it is significant to note the concentration of three extreme geomagnetic storms in 1941, whose intensities ranked fourth, twelfth, and fifth within the *aa* index between 1868 – 2010. Among them, the geomagnetic storm on 1 March 1941 was so intense that three of the four Dst station magnetograms went off scale. Herein, we reconstruct its time series and measure the storm intensity with an alternative Dst estimate (Dst*). The source solar eruption at 09:29 – 09:38 GMT on 28 February was located at RGO AR 13814 and its significant intensity is confirmed by large magnetic crochets of 35 nT measured at Abinger. This solar eruption most likely released a fast interplanetary coronal mass ejection with estimated speed 2260 km/s. After its impact at 03:57 – 03:59 GMT on 1 March, an extreme magnetic storm was recorded worldwide. Comparative analyses on the contemporary magnetograms show the storm peak intensity of minimum Dst* ≤ −464 nT at 16 GMT, comparable to the most and the second most extreme magnetic storms within the standard Dst index since 1957. This storm triggered significant low-latitude aurorae in the East Asian sector and their equatorward boundary has been reconstructed as 38.5° in invariant latitude. This result agrees with British magnetograms which indicate auroral oval moving above Abinger at 53.0° in magnetic latitude. The storm amplitude was even more enhanced in equatorial stations and consequently casts caveats on their usage for measurements of the storm intensity in Dst estimates.

## 1. Introduction

Solar eruptions occasionally direct interplanetary coronal mass ejections (ICMEs) towards the Earth. When they have sufficient speed, mass, and southward interplanetary magnetic field (IMF), such geo-effective ICMEs cause serious geomagnetic storms and variations in the terrestrial magnetic field (Gonzalez *et al*., 1994; Daglis *et al*., 1999). Intensities of the subsequent geomagnetic storms have been usually measured with the negative peak in the disturbance storm time (Dst) index since the International Geophysical Year (IGY: 1957 – 1958) as a representative of the ring current intensity. The Dst index is based on the average of the four mid-latitude magnetic disturbances at four reference stations (*e.g*., Sugiura, 1964; Sugiura and Kamei, 1999; WDC for Geomagnetism at Kyoto *et al*., 2015). Partially due to their geo-effectiveness, such solar eruptions have been subjected to astronomical interests from the early phase of the modern astrophysical studies (*e.g*., Carrington, 1859; Hale, 1931; Cliver, 2006; Gopalswamy, 2016; Tsurutani *et al*., 2020).

Within this chronological coverage, it was the March 1989 storm that recorded the most extreme intensity (minimum Dst = −589 nT; WDC for geomagnetism at Kyoto *et al*., 2015). This storm dramatically extended the auroral oval equatorward and had serious economic impacts on modern society with its space weather hazards (*e.g*., Allen *et al*., 1989; Cid *et al*., 2014; Pulkkinen *et al*., 2017; Riley *et al*., 2018; Boteler, 2019). This case shows that such extreme geomagnetic storms are not only of academic interest, but also of societal interest due to our increasing reliance on modern technological infrastructure (*e.g.*, Pulkkinen *et al*., 2017; Riley *et al*., 2018). Such extreme storms have been studied in terms of their chronological distributions (Kilpua *et al*., 2015; Lefevre *et al*., 2016) and revealed apparent correlations between the storm intensity and equatorward boundary of the auroral oval (Yokoyama *et al*., 1998).

Statistical analyses of such extreme storms are challenging as they are rare (Yerasimov *et al*., 2013) and only five storms went below the threshold of Dst = −400 nT within the coverage of the Dst index routinely derived by WDC Kyoto, (Riley *et al*., 2018; Meng et al., 2019). Among them, only the March 1989 storm developed beyond the threshold of Dst = −500 nT and has been considered as a superstorm (*e.g.*, Boteler, 2019). Furthermore, reconstructions of historical superstorms have shown that geomagnetic superstorms such as the events occurring in September 1859, February 1872, and May 1921 set further benchmarks both in terms of storm intensities (minimum Dst ≤ −800 nT) and equatorward boundaries of the auroral oval (*e.g.*, Tsurutani *et al*., 2003; Silverman and Cliver, 2001; Green and Boardsen, 2006; Cliver and Dietrich, 2013; Lakhina and Tsurutani, 2018; Hayakawa *et al*., 2018a, 2018b, 2019a, 2020c; Love *et al*., 2019b; Blake *et al*., 2020). Their impacts on modern technological infrastructure on the ground and in space are estimated to be even more catastrophic than their historical impacts (Daglis, 2001; Baker *et al*., 2008; Cannon *et al*., 2013; Hapgood, 2017; Riley *et al*., 2018; Oliveira *et al*., 2020).

Despite existing efforts on the extension of the Dst index (*e.g.*, Karinen and Mursula, 2005; Mursula *et al*., 2008), it is challenging to quantitatively measure the intensity of geomagnetic superstorms before the IGY, as their significant intensity frequently makes contemporary magnetograms run off scale (Riley, 2017, p. 118) and geographical coverage of historical magnetograms was much scarcer in the past. As such, Dst estimates (Dst*) for several superstorms have been reconstructed with alternative magnetograms in mid-latitude, with reasonable longitudinal separation, and quasi-completeness in the hourly resolution. These case studies have been conducted on four geomagnetic superstorms in October/November 1903 (minimum Dst* ≈ −513 nT; Hayakawa *et al*., 2020a), September 1909 (minimum Dst* ≈ −595 nT; Hayakawa *et al*., 2019b; Love *et al*., 2019a), May 1921 (minimum Dst* ≈ −907 ± 132 nT; Love *et al*., 2019b), and March 1946 (minimum Dst* ≤ −512 nT; Hayakawa *et al*., 2020b).

However, historical magnetograms show that these storms are only a small sample of extreme geomagnetic storms in history. Among them, the geomagnetic conditions in 1941 were especially notable, hosting at least three geomagnetic storms in March, July, and September, which ranked fourth, twelfth, and fifth within the *aa* index in 1868 – 2010, respectively (Lefèvre *et al*., 2016). Colourful auroral episodes for the September 1941 storm have been introduced in Love and Coïsson (2016), whereas little is known for the other two. On the other hand, the March 1941 storm had equally or even more extreme episodes, with magnetograms at three of four Dst stations going off-scale. In fact, this storm had the third largest Δ*H* deviation measured at the Greenwich-Abinger observatories (1650 nT to 1710 nT) within the 112 great storms during 1874 – 1954 (Jones, 1955, p.

79; see also Newton, 1941) and was ranked the fourth most intense within the *aa* index during 1868 – 2010 (Lefèvre *et al*., 2016). Shifting down to the mid-latitudes, this storm ranked in the seventh largest (> 560 nT) within the storms recorded at Kakioka Δ*H* between 1924 and 2020[1], despite its incomplete measurement. Therefore, in this manuscript, we analyse the time series of this extreme space/geomagnetic storm on 1 March 1941 with its source flare on 28 February 1941 from its source solar eruption to its terrestrial impact. In addition, we measure its intensity in Dst* based on alternative magnetograms and examine the equatorial boundary of the auroral oval.

## 2. Solar Eruption

In the declining phase of Solar Cycle 17 after its maximum in April 1937 (Table 1 of Hathaway, 2015; Figure 2 of Clette and Lefèvre, 2016), the solar surface in late February to early March 1941 was moderately eruptive but not as much as in April – September. Of the 27 solar flares recorded in the Hα observations, 7 flares were observed in RGO AR 13814 (D'Azambuja, 1942). Those flares were mostly categorised as 1 (Hα flare area = 100 – 250 msh) to 2 (Hα flare area = 250 – 600 msh) in their Hα flare area, whereas one on 3 March achieved importance of 3 (Hα flare area = 600 – 1200 msh) (D'Azambuja, 1942; see also Švestka, 1976, p. 14).

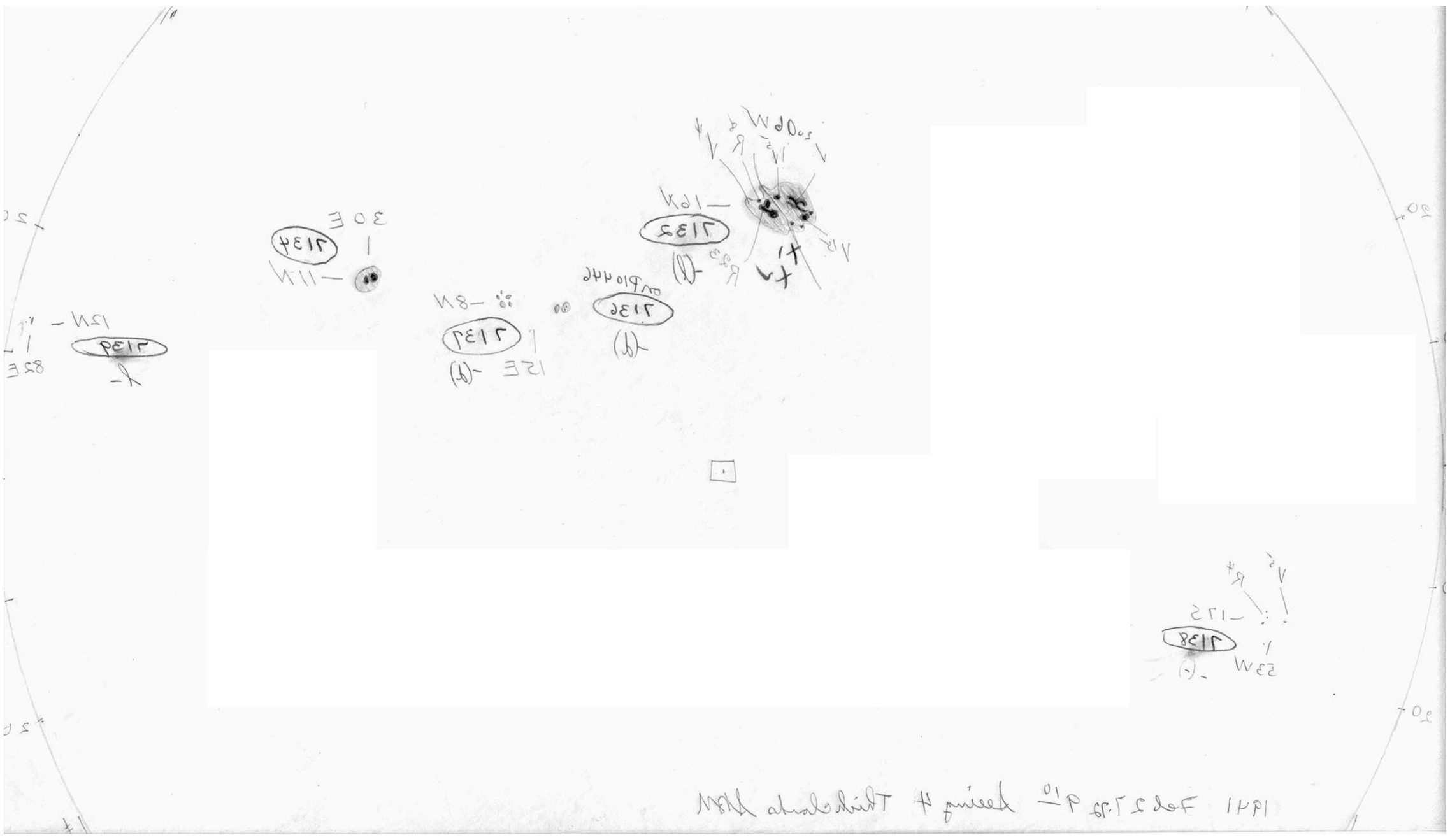

Figure 1: Sunspot drawing at Mt Wilson on 27 February 1941 with its side corrected, as viewed in

[1] https://www.kakioka-jma.go.jp/obsdata/Geomagnetic_Events/Events_index.php

the sky. The MWO AR 7132 corresponds to RGO AR 13814 (courtesy of the Mt. Wilson Observatory; see also Pevtsov *et al.*, 2019).

Upon contemporary observations, Newton (1941, p. 84) apparently associated our great storm with "a fairly large sunspot not far from the central meridian". As shown in Figure 1, Newton (1941) most likely indicated RGO AR 13814 with this description. Rather than the chromospheric eruptions, Newton (1941) listed three notable ionospheric disturbances with "wireless fade-outs" as footprints of solar flares at 15:45 – 16:10 GMT on 27 February, and 09:30 – 10:30 GMT and 15:27 – 15:40 GMT on 28 February. Among them, Newton (1941, p. 84) considered the second wireless fade-out as the source eruption and clarified the sky over Greenwich overcast. The third fade-out coincides chronologically with "a very small sudden commencement disturbance" at 15:26 on 28 February reported at Hermanus (Ogg, 1941, p. 372). This sunspot developed to 650 millionth solar hemisphere immediately after its central meridian passage on 27.3 February 1941 (Newton, 1941, p. 85).

Contemporary magnetograms confirm his discussions with magnetic crochets (see McIntosh, 1951), as footprints of intense X-ray radiation of solar flares (see Curto *et al.*, 2016; Curto, 2020). Figure 2 shows digitised magnetograms at Abinger, Eskdalemuir, and Lerwick on 28 February – 1 March 1941. We have digitised their scans hosted by BGS[2] based on the scale units shown in Hartnell (1922, p. 2) and the min-max scale values given in the observatory yearbooks (AMMO, 1958, p. 21, 109; Jones, 1954, p. D21). Our digitisations have located magnetic crochets around 09:29 – 09:38 GMT on 28 February with notable amplitudes: ≈ 13 nT at Lerwick, ≈ 20 nT, at Eskdalemuir, and ≈ 35 nT at Abinger, respectively (*c.f.*, McIntosh, 1951; Jones, 1955, p. 81).

Interestingly, significant polar cap absorption was reported at Tikhaya Bay (N80°19′, E52°47′) for 4 days since somewhere between 22 GMT on 26 February and 6 GMT on 27 February and caused blackouts for 99 hours (Besprozvannaya, 1962, p. 147; see also Cliver et al., 1990, p. 17109). Such polar cap absorptions indicate presence of solar proton events (*e.g.*, Shea and Smart, 2012), as shown from their known pairs with the intense solar proton events (Besprozvannaya, 1962; McCracken, 2007; Usoskin *et al.*, 2020). While only one minor flare was reported in this onset interval (0:36 UT on 27 February at Mt. Wilson; D'Azambuja, 1942, p. 80), this implies some of the flares in this sequence from RGO AR 13814 probably caused notable solar proton events. This implies that the source flare around 09:29 – 09:38 GMT on 28 February also possibly caused a solar proton event.

[2] https://www.bgs.ac.uk/data/Magnetograms/home.html

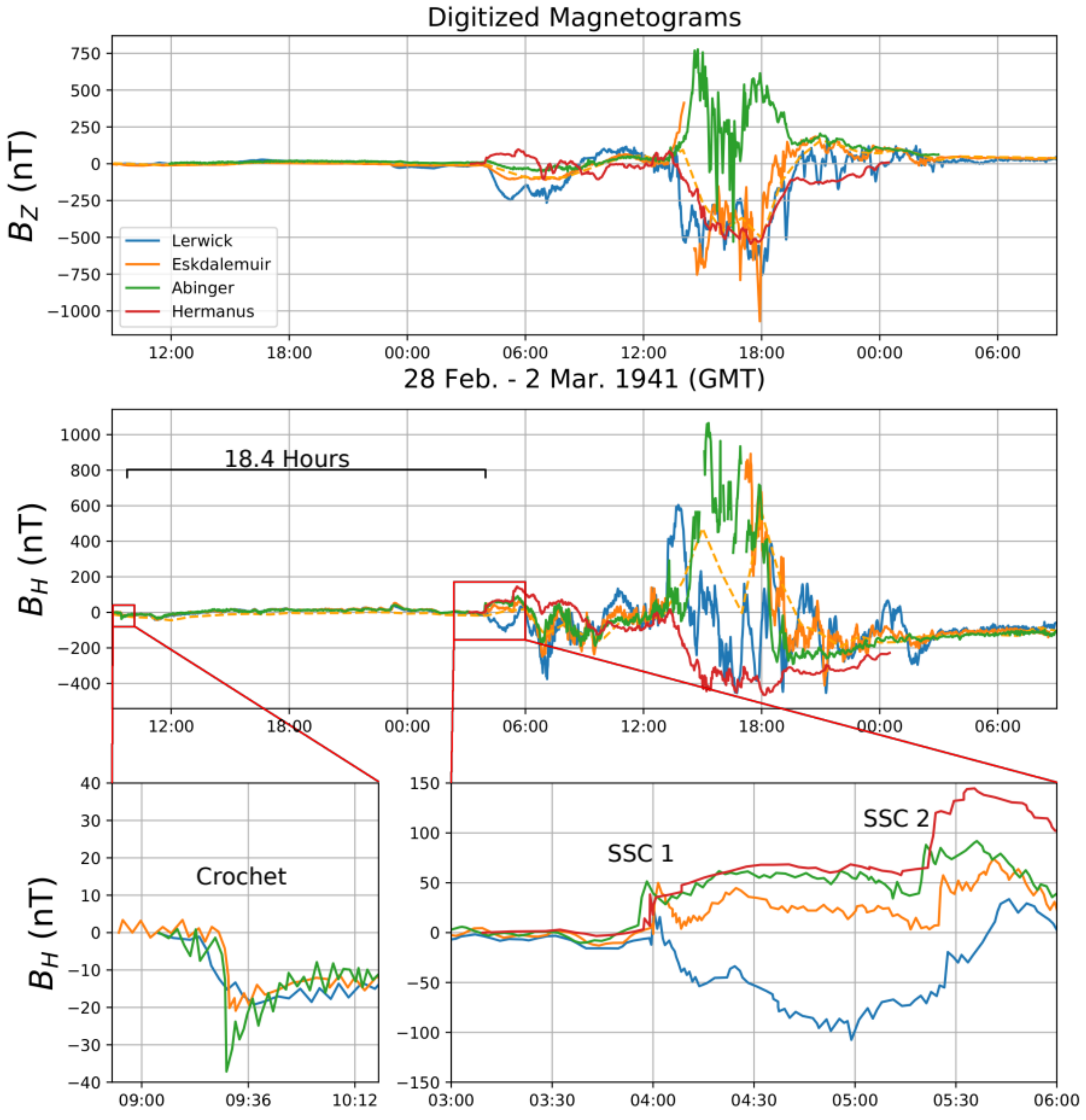

Figure 2: Magnetic disturbances at Lerwick, Eskdalemuir, and Abinger spanning from 28 February to 1 March 1941 digitised from their original magnetograms at BGS and trace copy of Ogg (1941). The dashed orange lines show the recorded hourly values at Eskdalemuir (AMMO, 1958). These magnetograms show magnetic crochets at 09:29 – 09:38 GMT and the SSCs at 03:57 – 03:59 GMT.

## 3. Interplanetary Coronal Mass Ejection

These crochets were followed by storm sudden commencements at ≈ 03:57 – 03:59 on 1 March, as summarised in Table 1. Among them, Hermanus reported two SSCs at 03:58 ($\Delta H$ = 42 nT) and 05:19 ($\Delta H$ = 48 nT) which probably indicates arrivals of at least two consecutive ICMEs. The time lag between the magnetic crochet and the SSCs shows the transit time of the ICME as 18.4 hours and imply the average speed of this ICME as 2260 km/s.

Siscoe et al. (1968) and Burton et al. (1975) provide the following empirical formula for dynamic pressure computations as a function of geomagnetic field variations:

$$\Delta P_d[\text{nPa}] = 4.0\times 10^{-3} Dst^2 \ . \quad (1)$$

At the instance of maximum compression, Dst* shows the SSC amplitude $\Delta$B ~ 35 nT, and equation (1) yields $\Delta P_d$ ~ 4.9 nPa. This ram pressure is almost 8 times larger than the ram pressure during quiet solar wind conditions, namely ~0.75 nPa for solar wind speed of 300 km/s and solar wind density of 5 $cm^{-3}$. In addition, by assuming equilibrium conditions during the interactions of the magnetosphere with the solar wind, the resulting inward magnetopause position is given by (Baumjohann and Treumann, 2009):

$$X_{mp} = \left(\frac{K B_0{}^2}{2\mu_0 P_d}\right)^{1/6}, \quad (2)$$

where $B_0$ is the contemporary dipole magnetic field at the Earth's surface at $L$=1, $\mu_0$ is the vacuum permeability, and $K$ is an arbitrary factor. According to the IGRF model for the epoch of 1940 (Thébault *et al.*, 2015), the spherical harmonic coefficients $g_1^0$, $g_1^1$, and $h_1^1$ are –30654 nT, –2292 nT, and 5821 nT, respectively, which yield $B_0$ of 32286 nT. Assuming that $K$ = 2 and that $P_d$ = 5.65 nPa (=0.75+4.9 nPa), $X_{mp}$ ~ 7.2 $R_E$. By comparing this result with the magnetopause standoff position during nominal solar wind conditions (~10.1$R_E$ for $P_d$ = 0.75 nPa), one concludes that the magnetopause moved ~3$R_E$ inwards. However, the magnetopause did not surpass the threshold of geosynchronous orbit (~6.6 $R_E$), which could expose satellites to the hostile solar wind environment (*e.g.*, Baker *et al.*, 2017). The most inward observed magnetopause motion, ~5.24 $R_E$ from the Earth's center (Hoffman et al., 1975) occurred as a result of the impact of the fastest CME to hit Earth ever recorded, with speed 2850 km/s (Vaisberg and Zastenker, 1976; Cliver *et al.*, 1990). Therefore, these results show that the impact of the likely CME in March 1941 had moderate compression of the Earth's magnetosphere.

| Observatory | lat. | long. | mlat. | mlon. | SSC | H-range | References |
|---|---|---|---|---|---|---|---|
| Heramnus* | S34°25′ | E019°13′ | -33.2 | 80 | 42&48 | 614 | Ogg 1941 |
| Watheroo* | S30°19′ | E115°52′ | -41.8 | -174.9 | 19 | 658 | Parkinson 1941 |
| Tucson* | N32°15′ | W110°50′ | 40.4 | -48.4 | 74 | > 550 | White 1941 |
| Apia* | S13°48′ | W171°45′ | -16.2 | -100.2 | | | WDC Kyoto |
| Kakioka | N36°14′ | E140°11′ | 26 | -154.5 | 31 | > 560 | KED |
| Alibag | N18°38′ | E072°52′ | 9.5 | 143.2 | 42 | > 785 | Rangaswami 1941 |
| Huancayo | S12°02′ | W076°20′ | -0.6 | -7.6 | 80 | 1180 | Ledig 1941 |
| Abinger | N50°11′ | E000°23′ | 53 | 83.1 | 50 | 1650 | Figure 2 |
| Eskdalemuir | N55°19′ | W003°12′ | 58.6 | 82.5 | 55 | 1320 | Figure 2 |
| Lerwick | N60°08′ | W001°11′ | 62.6 | 88.2 | 40 | 1180 | Figure 2 |

Table 1: Recorded amplitudes of SSC and geomagnetic storms ($\Delta H$) at each observatory. Their magnetic latitude is computed with IGRF-12 model (Thébault *et al.*, 2015). The stations with asterisk (*) are the reference stations in our estimate. The observational data at Apia are acquired only with its hourly value from the WDC Kyoto (WDC for geomagnetism at Kyoto) and do not show the SSC amplitude and the H-variation in the spot value.

**4. Magnetic Disturbance**

After the ICME arrival at 03:57 GMT, the intense magnetic storm developed so rapidly that a number of the ground-based magnetograms went off the scale at that time. This makes the estimate of the intensity for this storm rather challenging. While the Dst index has been used to evaluate storm intensity as a quantitative measurement for the ring current development, the March 1941 storm occurred far before its introduction. The standard Dst index has been measured with an average of the four stations (Kakioka, Honolulu, San Juan, and Hermanus) with weighting of their magnetic latitude (Sugiura, 1964; WDC for Geomagnetism at Kyoto *et al.*, 2015). However, the March 1941 storm was only incompletely recorded at three of these standard stations, during 14 – 16 and 18 GMT at Kakioka, 16 – 17 GMT at Honolulu, and 14 – 22 GMT at San Juan[3]. These data gaps affect any attempts to extend the Dst index upon this storm with the use of existing Dst reference stations (see Riley, 2018; *c.f.*, Karinen and Mursula, 2005, 2006).

As such, we need to substitute these three stations with three mid-to low-latitude magnetograms in similar magnetic longitudes without significant data gaps to estimate Dst equivalent measure of the storm intensity. Three magnetograms at Watheroo (WAT), Apia (API), and Tucson (TUC) satisfy this requirement and are used to replace the incomplete data from Kakioka, Honolulu, and San Juan, respectively. We have derived their hourly data from the WDC for geomagnetism at Kyoto. Following the procedure of the standard Dst calculations, we have first derived the hourly disturbance at each station ($D_i$ (t)), subtracting the baseline ($B_i$) and solar quiet field variations ($Sq_i$ (t)). We then averaged the hourly disturbance at each station with weighting of their contemporary magnetic latitude ($\lambda_i$) (Sugiura, 1964; WDC for geomagnetism at Kyoto *et al.*, 2015). These calculations are summarised by the following two equations.

[3] http://wdc.kugi.kyoto-u.ac.jp/index.html

$$D_i(t) = H_i(t) - B_i - Sq_i(t)\,, \tag{3}$$

$$Dst(t) = \frac{1}{4}\sum_{i=4}^{4}\frac{D_i(t)}{\cos(\lambda_i)} \tag{4}$$

In our analyses, the baseline ($B_i$) has been approximated with the observatory annual means for each station, provided in the WDC for geomagnetism at Edinburgh[4]. The solar quiet field variation ($Sq_i$ (t)) has been approximated with the five quietest days in the previous month of this storm (27, 12, 19, 18, and 1 February 1941), which have been selected on the basis of the revised daily *aa* index provided in Lockwood *et al.* (2018a, 2018b). The magnetic latitude of each station in 1941 has been computed with the angular distance of these stations and position of the magnetic pole in 1941 using the archaeo-magnetic field model IGRF-12 (Thébault *et al.*, 2015).

Its validity has been checked in comparison with the standard Dst index provided in WDC for Geomagnetism at Kyoto *et al.* (2015). As the hourly data of Watheroo Observatory is available only until 1958 (WDC for Geomagnetism at Kyoto, 2020), we have only a two-year overlap with the standard Dst index. Here, we have examined the extreme geomagnetic storm in 1957 September, the largest storm in this interval, and the third-largest in the coverage of the standard Dst index (WDC for geomagnetism at Kyoto *et al.*, 2015). With our procedure and the same selections of the reference stations (WAT, HER, TUC, and API),, we have computed its minimum Dst* estimate ≈ −399 nT, in contrast with that of = −427 nT in the official Dst index. This shows a difference of 28 nT and a relative difference of 7.6%, which is almost consistent with the variation of Dst estimate and standard Dst index derived in Love *et al.* (2019). Note that both of the test cases in our study and Love *et al.* (2019) show the Dst estimate as more conservative than the standard Dst value. As such,

[4] http://www.geomag.bgs.ac.uk/data_service/data/annual_means.shtml

this result confirms the validity of our method with equivalent reference stations.

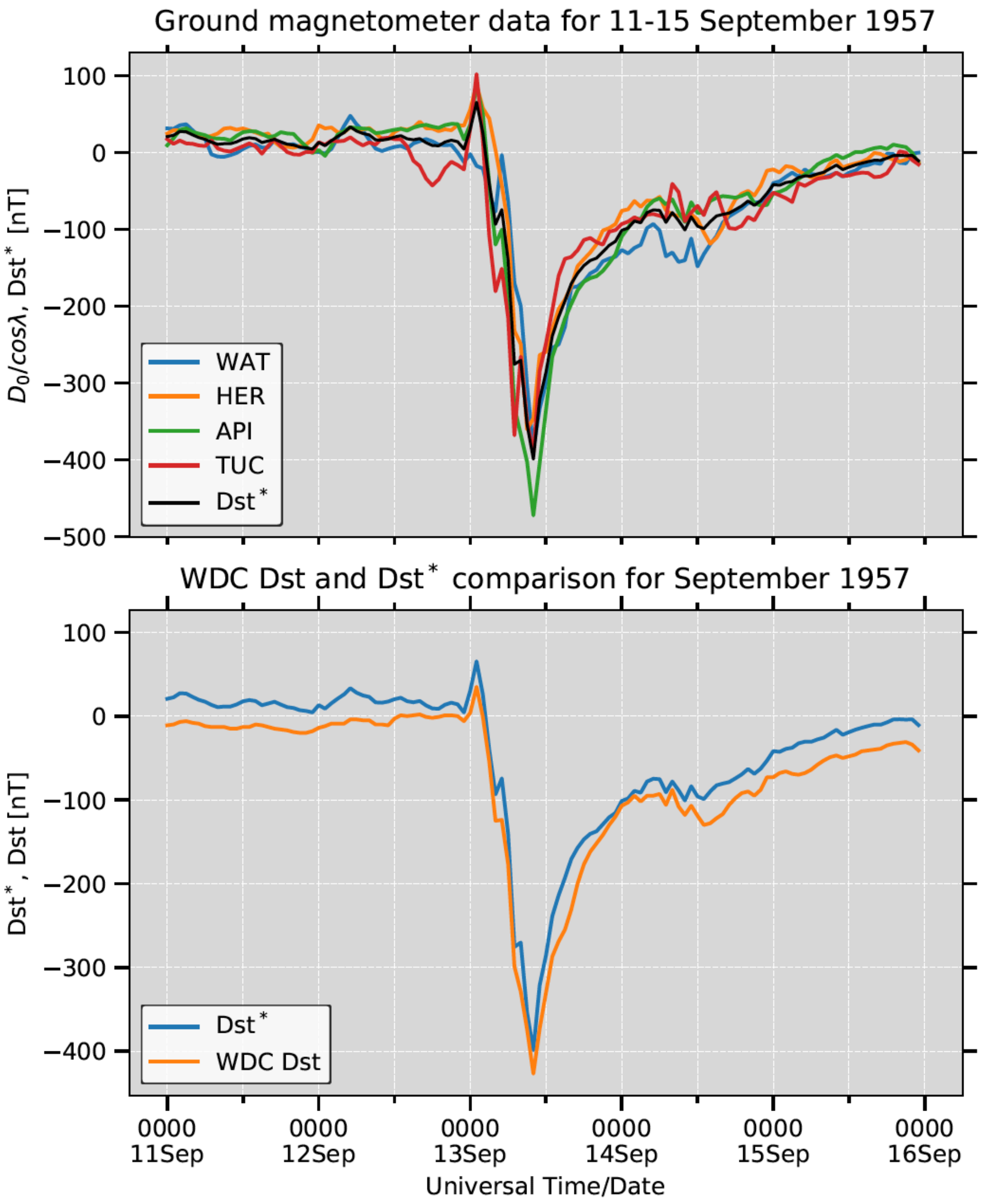

Figure 3: Comparison of the standard Dst index and our Dst* estimate for the September 1957 storm. For the Dst* estimate, we have used WAT, HER, TUC, and API.

Our Dst* estimate and the hourly disturbance at each observatory with latitudinal weighting have been summarised in Figure 4. Our reconstruction shows its intensity in the minimum Dst* estimate of ≈ −464 nT. The reconstructed time series shows a positive excursion peaked at 79 nT at 05 GMT after the ICME arrival at 03:57 GMT, probably because of the combination with the second SSC at 05:09 GMT reported at Hermanus (Ogg, 1941). Afterward, this storm shows an initial negative excursion down to −180 nT at 09 GMT with temporal recovery to −64 nT at 12 GMT. The magnetic field was exposed to a steep decrease at 15 GMT and then peaked to −464 nT at 16 GMT on 1 March and a gradual recovery phase up to early 3 March. This "two-step" time series is consistent with the frequent geomagnetic behaviour in major geomagnetic storms (Kamide *et al.*, 1998), resulted from variable IMF, a combination of shock-sheath and following magnetic cloud, or a combination of multiple ICMEs (Daglis *et al.*, 2003; Richardson and Zhang, 2008; Lugaz et al., 2016). Near the first negative dip, the $D_0/\cos\lambda$ value decreased steeper at API than at HER. This can be attributed to the asymmetric development of the storm-time ring current (Cummings, 1966), in which the intensity of the storm-time ring current is highest in the dusk-midnight sector, whereas it is lowest in dawn-noon sector. The in-situ observation shows that the asymmetry is prominent during the storm main phase because fresh ions are transported from the nightside plasma sheet to the duskside by the enhanced magnetospheric convection (Ebihara *et al.*, 2002). During the recovery phase, the asymmetry is relaxed because of the dominance of the westward grad-B and curvature drift motion of the ions (Ebihara *et al.*, 2002). Note that the API magnetogram has a data gap on 2 March in UT, whereas this data gap falls in the gradual recovery phase and hence does not affect our intensity reconstruction. In fact, our Dst* estimate curve (continuous black curve) and one without API data (broken black curve) are shown in Figure 4. Closer inspection shows that, at Tucson, "At about 16 h the *H*-reserve spot went off scale negative for a range in excess of 550 gammas" (White, 1941, p. 257), whereas its hourly data are without break. Therefore, it is probable that Tucson magnetogram went off scale for a short time and made the hourly value slightly more conservative than the reality. Therefore, we consider our intensity estimate as a conservative estimate and describe its intensity as minimum Dst* ≤ −464 nT.

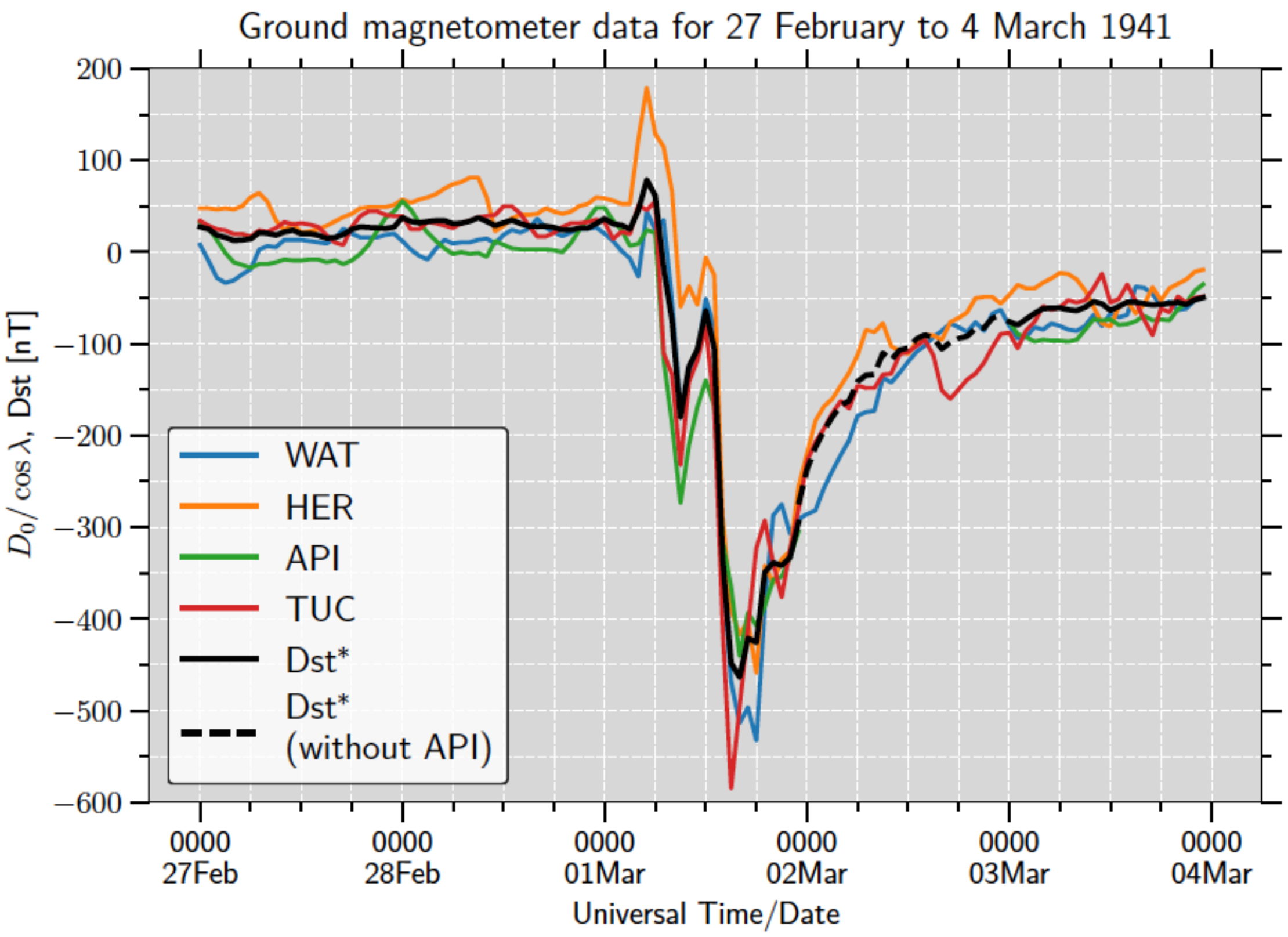

Figure 4: the Dst* estimate and the individual disturbance time series at each reference station for 27 February to 4 March 1941. The Dst* estimate is shown in a continuous black curve (except for 3 March) and a broken black curve on 3 March without API data. The corrected time series of WAT, HER, API, and TUC are shown in blue, orange, green, and red, respectively.

**5. Low-Latitude Aurorae**

Around its peak recorded at 16 GMT on 1 March, the East Asian sector was favourably situated in the midnight sector for the auroral visibility. Indeed, the aurorae have been reported in Manchuria and northern Japan at that time. In Manchuria, aurorae were reported at Hăilāĕr (N49°13′, E119°46′; 37.8° MLAT) during 23:05 (1 March) – 00:18 (2 March) in LT and 01:40 – 02:20 (2 March) in LT, namely 15:05 – 16:18 and 17:40 – 18:20 (1 March) in GMT. The aurorae were reported as reddish glows moving from NW to E in the first part and then as reddish glows with stripes moving from NE to NW in the second part (*Shèngjīng Shíbào*, 1941-03-08, p. 4). This implies that the aurora moved eastward in the premidnight, and westward in the postmidnight. Its motion is, in part, consistent with low-latitude aurorae, in which the eastward motion and the westward motion were observed in the dusk-side (Shiokawa *et al*., 1994).

In northern Japan, aurorae were reported on both sides of Soya Strait. At Otomari (N46°38′, E142°46′; 36.5° MLAT), the aurorae were reported during 23:00 (1 March) – 03:55 (2 March) in LT (14:00 – 18:55 on 1 March in GMT), as reddish glow with various intensities. At Wakkanai (N45°25′, E141°40′; 35.2° MLAT), diffuse reddish aurorae were visible during 22:56 (1 March) – 04:30 (2 March) in LT (13:56 – 19:30 on 1 March in GMT). At its maximum, the aurora altitude reached almost up to the zenith. At Oshitomari (N45°14′, E141°13′; 35.0° MLAT), red-yellowish glow with stripes in bluish white was visible during 01:05 – 03:30 (2 March) in LT (16:05 – 18:30 on 1 March in GMT) and its altitude reached up to 60 – 70° at its maximum (Kisho Yoran, 1941, pp. 267 – 269; see Figure 5).

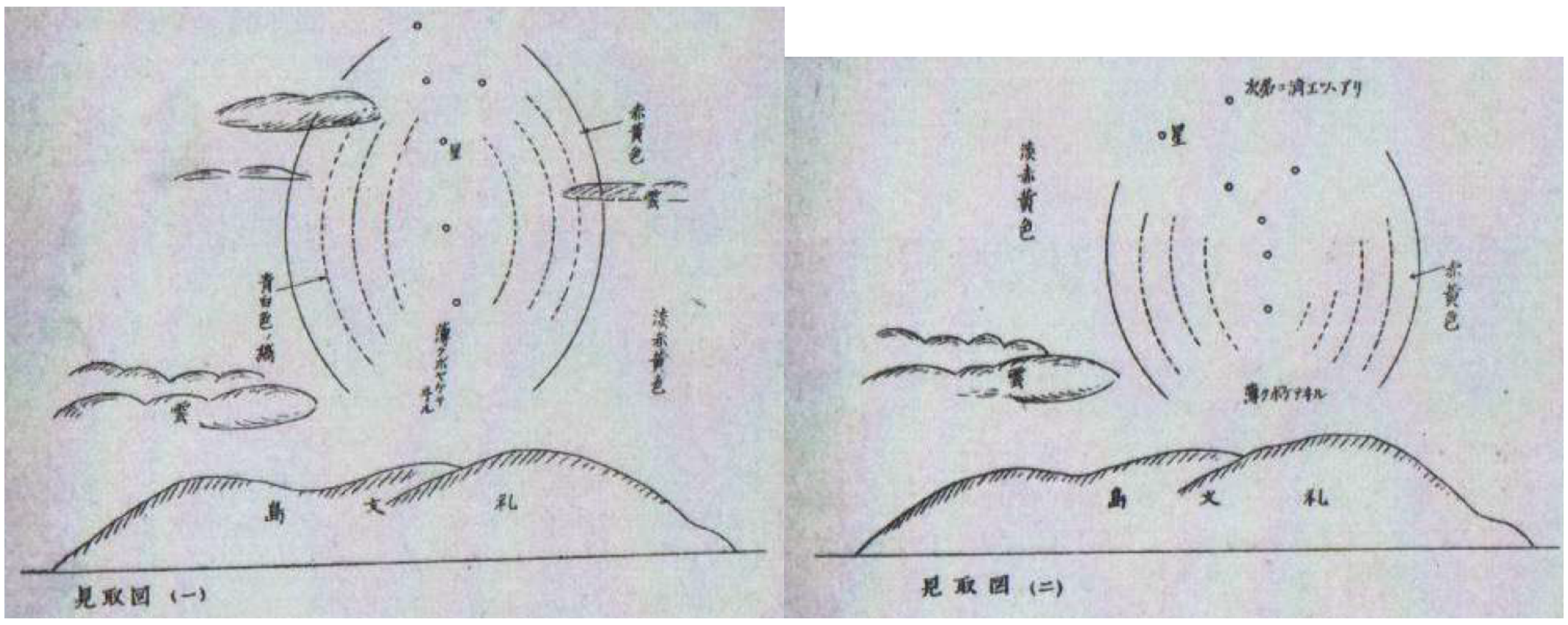

Figure 5: Japanese auroral sketches at Oshitomari reproduced from Kisho Yoran (1941, p. 268) in red-yellowish background with bluish-white stripes at its maximum (left) and after its decay (right). The bluish-white stripes are shown in continuous lines and broken lines. The red-yellowish glows are shown in the background sky. The island depicted below is Rebun Island northward from Oshitomari.

The auroral visibility in the East Asian sector lasted ≈ 14 – 18.5 GMT and is chronologically consistent with the main phase of this geomagnetic storm. In addition, a short telegraph disturbance was reported at Oshitomari during 01:55 – 02:05 on 2 March in LT (16:55 – 17:05 GMT on 1 March) and this is chronologically located immediately after the peak of this storm at 16 GMT on 1 March (Kisho Yoran, 1941, pp. 267 – 269).

Among these records, those at Wakkanai at 35.2° MLAT and Oshidomari at 35.0° MLAT provide the elevation angle of the auroral visibility almost up to the zenith and up to 60 – 70°, respectively. On their basis, the equatorward boundary of the auroral oval (EBAO) is estimated 37.6° and 38.5° in invariant latitude (ILAT) in terms of footprints of the magnetic field lines, assuming the auroral altitude as ≈ 400 km (Roach *et al*., 1960; Ebihara *et al*., 2017). These values almost coincide with one another. The slight equatorward extension in the former may be tentatively associated with the different extension of the SAR (stable auroral red) arcs and reddish aurorae. This is because SAR arcs are typically without visible structure (Kozyra *et al*., 1997) and consistent with the Wakkanai report with diffuse reddish glow. On the other hand, what was described in the Oshitomari report is certainly an auroral display, being described as a reddish glow with stripes respectively (Kisho Yoran, 1941, pp. 267 – 269).

**6. Conclusion and Discussion**

In this contribution, we have analysed the time series of the extreme space weather event in February – March 1941. Despite the lack of optical evidence of the source solar flare, contemporary magnetograms have shown significant magnetic crochets ($\Delta H \approx 35$ nT at Abinger) at 09:29 – 09:38 GMT on 28 March. This is synchronised with a significant polar cap absorption lasting for 4 days since 26/27 February and probably associated with a notable solar proton event.

The driving ICME impact was recorded with SSCs at 03:57 – 03:59 on 1 March (Table 1). On their basis, the ICME speed has been computed as 2260 km/s. By applying well-known empirical models for the solar wind dynamic pressure and the magnetopause stand-off position at the instance of maximum compression, it was found that they are 4.9 nPa and 7.2 $R_E$, respectively. This calculated average ICME speed ranks the ICME as being the fourth fastest on record following those in August 1972, September 1859, and February 1946 (Cliver *et al*., 1990; Cliver and Svalgaard, 2004; Gopalswamy *et al*., 2005; Knipp *et al*., 2018; Chertok, 2020). This subsequent geomagnetic activity may have been caused by the CME sheath and/or magnetic cloud (Lugaz *et al*., 2016; Kilpua *et al*., 2019).

We have reconstructed its time series and intensity (minimum Dst* ≤ −464 nT at 16 GMT on 1 March 1941) based on the hourly magnetic measurements at Hermanus, Watheroo, Apia, and Tucson. Our intensity estimate has placed this storm in an extreme category, as only five geomagnetic storms exceeded the threshold of Dst = −400 nT during the space age, for the last 63 years (*e.g.*, Meng *et al*., 2019). Indeed, this intensity is more extreme than the second most intense storm in July 1959 (minimum Dst = −429 nT) but considerably less than the most intense storm in March 1989 (minimum Dst = −589 nT), within the standard Dst index since the IGY (WDC for geomagnetism at Kyoto *et al.*, 2015). While this storm is slightly more moderate than historical superstorms in September 1859, February 1872, October/November 1903, September 1909, May 1921, and March 1946 (Cliver and Dietrich, 2013; Hayakawa *et al.*, 2018a, 2019a, 2019b, 2020a, 2020b, 2020c; Love *et al.*, 2019a, 2019b), its intensity is significantly notable and in modern times its consequences would be quite serious.

With its peak around 16 GMT, the East Asian sector was favourably situated for the auroral observations. Indeed, the East Asian auroral records imply the EBAO as 38.5° ILAT. The temporal and spatial evolution of the auroral oval compares well with the magnetic disturbances in the high-resolution British magnetograms (Figure 2). From the first SSC at 03:58, the vertical components ($B_z$) at each of the British operated observatories appear to be broadly coherent. From 13:30 − 14:30, Lerwick witnessed a sharp decrease in $B_z$ with a simultaneous increase in the horizontal component ($B_H$), while Abinger and Eskdalemuir witnessed a large increase $B_z$, and a staggered increase in $B_H$. This most likely indicates that an eastward Hall current flew broadly in the ionosphere over Lerwick (62.6° MLAT) and Eskdalemuir (58.6° MLAT) (Rostocker, 1973). After 14:30, the broadly coherent variations between the Eskdalemuir and Lerwick B*z* suggests that the eastward Hall current had moved southward of Eskdalemuir. The minimum reported Bz for Abinger in Jones (1954) may indicate that the eastward ionospheric current went southward beyond Abinger (53.0° MLAT) by 16:30. Around this time, Abinger saw large positive $B_H$ variations, indicating the proximity of auroral currents. From 17:10 onwards, the $B_z$ at Abinger is anti-correlated with Lerwick and Eskdalemuir, again indicating an eastward ionospheric current between Eskdalemuir and Abinger, until around 20:30, when the geomagnetic activity begins to gradually subside at all sites.

On the other hand, Hermanus shows characteristics of the ring current instead of ionospheric currents. This indicates that Lerwick was at one point probably near the poleward boundary of the auroral oval, Eskdalemuir and Abinger were almost under the auroral oval except for the storm maximum, and the auroral oval probably extended more equatorward than Abinger around the maximum. The auroral oval likely did not reach latitudes low enough to affect measurements at Hermanus in the Southern Hemisphere. Assuming that the evolution of the auroral ovals in the both hemispheres were similar with one another, this time series agrees with the maximum of the auroral activity around 16 GMT and its spatial evolution of |38.5|° ILAT, situated between MLATs of Abinger (|53.0|° MLAT) and Hermanus (|33.2|° MLAT).

This temporal evolution compares the auroral expansion with those during the October/November 1903 storm (minimum Dst* ≈ −531 nT vs. EBAO ≈ 44.1° ILAT; Hayakawa *et al*., 2020a), the March 1946 storm (minimum Dst* ≤ −512 nT vs. EBAO ≤ 41.8° ILAT; Hayakawa *et al*., 2020b), and the March 1989 storm (minimum Dst = −589 nT vs. EBAO = 35 – 40.1° ILAT; Rich and Denig, 1992; Boteler, 2019). As such, this auroral expansion is certainly not as extreme as those during the superstorms in 1859 and 1921 (minimum Dst* ≈ −900 nT; Cliver and Dietrich, 2013; Love *et al*., 2019; Hayakawa *et al*., 2019a) but still are comparable with those around the minimum Dst* ≈ −500 nT.

Interestingly, as shown in Table 1, the reported amplitudes of geomagnetic disturbance in spot value are more significant at the equatorial stations (1180 nT at Huancayo and > 785 nT at Alibag) than at mid-latitude stations (> 550 nT at Tucson and 614 nT at Hermanus), while the lost value with the magnetogram saturations still reserve possibility of local extreme disturbances at mid-latitude stations. This is comparable with the trend seen in the March 1946 storm, where the equatorial magnetograms showed more significant disturbances than the mid-latitude magnetograms (Table 1 of Hayakawa *et al*., 2020b). The contrast of the minimum Dst* ≤ −464 nT with Huancayo spot value = 1180 nT casts serious caveats on existing discussions of the 1859 geomagnetic superstorms with spot value at the single equatorial station (*e.g.*, Tsurutani *et al*., 2003), and requires investigations of further mid-latitude magnetograms to improve its intensity estimate.

**Acknowledgment**
This work was supported in part by JSPS Grant-in-Aids JP15H05812 and JP17J06954, JSPS Overseas Challenge Program for Young Researchers, the 2020 YLC collaborating research fund, and the research grants for Mission Research on Sustainable Humanosphere from Research Institute for Sustainable Humanosphere (RISH) of Kyoto University and Young Leader Cultivation (YLC) program of Nagoya University. We thank Mt. Wilson Observatory for providing sunspot drawings on 27 February 1941, WDC for Geomagnetism at Edinburgh for providing geomagnetic baselines and British magnetograms, WDC for Geomagnetism at Kyoto for providing the Dst index and magnetic measurements at Hermanus, San Juan, Honolulu, and Watheroo, Kakioka Event Database for providing data on the SSC and magnetic storms observed in the said observatory, WDC SILSO for providing international sunspot numbers, and Solar Science Observatory of the NAOJ for providing copies of *Quarterly Bulletin on Solar Activity*. Magnetograms were digitised using the WebPlotDigitizer software (https://automeris.io/WebPlotDigitizer/). SPB was supported by the NASA's Living With a Star program (17-LWS17_2-0042). HH thanks Margaret A. Shea for her helpful advices on the historical solar proton events.